\documentclass{elsart}
\usepackage{epsfig}
\def\ccbar{\mbox{c}\overline{\mbox c}}
\def\qqbar{\mbox{q}\overline{\mbox q}}
\def\ee{\mbox{e}^+\mbox{e}^-}
\def\tot{{\rm tot}}
\def\tag{{\rm tag}}
\def\bg{{\rm bg}}
\def\s{{\rm s}}
\begin{document}

\begin{frontmatter}

\title{Classifying LEP Data with Support Vector Algorithms}
\thanks[talk]{To be submitted to the Proceedings of
AIHENP99, Crete, April 1999}
\author[opal]{P. Vannerem}
\author[gmd]{K.-R. M\"uller}
\author[gmd]{B. Sch\"olkopf}
\author[gmd]{A. Smola}
\author[opal]{S. S\"oldner-Rembold}
\address[opal]{Fakult\"at f\"ur Physik, Hermann-Herder-Str. 3, 
D-79104 Freiburg, Germany}
\address[gmd]{GMD-FIRST, Rudower Chaussee 5, D-12489 Berlin, Germany}

\begin{abstract}
We have studied the application of different classification
algorithms in the analysis of simulated high energy physics data.
Whereas Neural Network algorithms have become a standard tool for 
data analysis, the performance of other classifiers such as Support 
Vector Machines has not yet been tested in this environment.\\
We chose two different problems to compare the performance of a 
Support Vector Machine and a Neural Net trained with back-propagation: 
tagging events of the type
$\ee\to\ccbar$ and the identification of muons produced in multihadronic 
$\ee$ annihilation events. 
\end{abstract}
\end{frontmatter}

\section{Classification algorithms}
Artificial Neural Networks (ANN) are a useful tool to solve
multi-dimensional classification problems
in high energy physics, in cases where one-dimensional cut techniques
are not sufficient.
They are used both as hard-coded chip for very fast 
low-level pattern recognition in on-line triggering~\cite{CDF,H1} and as a 
statistical tool for particle and event classifications in offline 
data analysis. 
In offline data analysis, a Monte Carlo simulation of the physics
process and the detector response is necessary to train an ANN by
supervised learning. 
ANN algorithms have been applied successfully in 
classification problems such as gluon-jet tagging~\cite{gluon} and 
b-quark tagging~\cite{btag}. The ANN classifiers~\cite{jetnet} constructed 
in this paper have sigmoid nodes. Design and tuning issues were solved
by applying practical experience rules~\cite{mueller1}.
\begin{figure}[bt]
\begin{center}
\epsfig{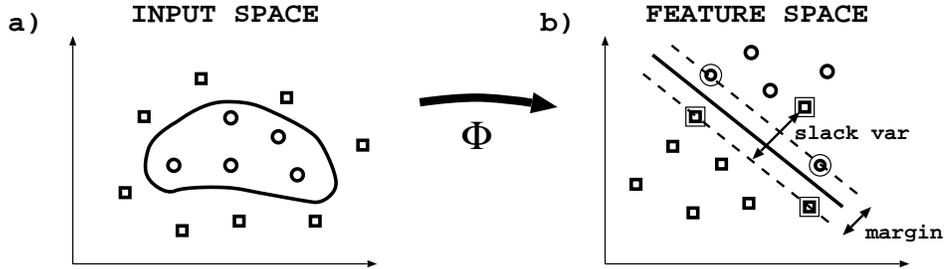}
\end{center}
\caption{A Support Vector Machine maps the input vectors of 2 different 
classes (a) into a higher dimensional space (b) and determines the 
separating hyperplane with maximum \emph{margin}
in that space. For non-separable problems additional \emph{slack variables}
are introduced.
}\label{fig:svm}
\end{figure}

This paper presents the application of a recently proposed 
machine-learning method, called Support Vector 
Machines (SVM)~\cite{vapnik1,vapnik2}, to high energy physics data.
The underlying idea is to map the patterns, i.e., the
$n$-dimensional vectors $\bf x$ of $n$ input variables, from the 
input space to a higher dimensional feature space with a
non-linear transformation (Fig.~\ref{fig:svm}). Gaussian radial basis 
functions are used as a kernel for the mapping.
After this mapping the problem becomes linearly separable by hyperplanes. 
The hyperplane which maximises the margin is defined by
the support vectors which are the patterns lying closest to the
hyperplane. This hyperplane which is determined with a training set
is expected to ensure an optimal separation of the different classes in 
the data.
 
In many problems a complete linear separation of the patterns is not 
possible and additional slack variables for patterns not lying on the 
correct side of the hyperplane are therefore introduced.
The training of a SVM~\cite{mueller2,mueller3} is a convex optimisation problem.
This guarantees that the global minimum can be found, 
which is not the case when minimising the error function for an ANN 
with back-propagation. The CPU time needed to find the hyperplane
scales approximately with the cube of the number of patterns.

\section{Data sets and the OPAL experiment}
Two distinctly different problems were chosen for comparing 
ANN and SVM classifiers, charm tagging and muon identification.

The first problem is to classify (tag) $\ee\to\qqbar$ events
according to the flavour of the produced quarks, 
separating c-quark events from light quark (uds) and b-quark
events. Flavour tagging is necessary for precision measurements
of electroweak parameters of the Standard Model.
The events are divided into two hemispheres by a plane perpendicular to the 
thrust axis. The flavour tag is applied separately to both hemispheres,
which contain the jets from the two produced quarks.
For a signal (s) with background (bg) the 
efficiency $\varepsilon$ is defined as 
$$\varepsilon = N^\s_{\tag}/N^\s_{\tot},$$
where $N^\s_{\tag}$ are the number of correctly tagged hemispheres
and $N^\s_{\tot}$ are all signal hemispheres in the sample. 
The purity $\pi$ is given by
$$\pi = N^\s_{\tag} / \left( N^\s_{\tag} + N^{\bg}_{\tag} \right)$$
with $N^{\bg}_{\tag}$ being the number of tagged background hemispheres. 

Due to the high mass ($\simeq 5$~GeV) and long lifetime ($\simeq 1.5$~ps) 
of b hadrons, hemispheres containing b-quarks can be tagged using an ANN
with typical efficiencies of 25\% and purities of about 92\%~\cite{opalRb}.
 
\begin{figure}[tb]
\begin{center}
\epsfig{file=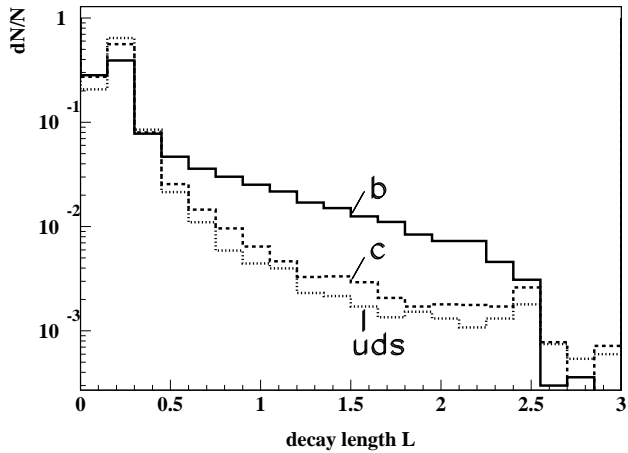,height=.5\textwidth}
\end{center}
\caption{The distribution of the reconstructed decay length $L$ is an input
distribution to the flavour tag. 
\label{fig:dcyl}}
\end{figure}

However, for the lighter charm quark, the measured fragmentation properties and
secondary vertex quantities are very similar in charm events
and uds events (Fig.~\ref{fig:dcyl}).
High purity charm tags with low efficiency are possible 
using D$^{*\pm}$ mesons
or leptons with high transverse momentum from semi-leptonic decays.
Applying an ANN or SVM charm tag to kinematic variables defined
for all charm events is expected to increase the charm tagging efficiency
at the cost of lower purities.

The second problem is the identification of muons which are produced
in the fragmentation of $\ee\to\qqbar$ events. Muons
are usually not absorbed in the calorimeters. They 
are measured in muon chambers which build the outer layer of a typical
collider detector. 
A signal in these chambers which is matched to a track in
the central tracking chamber is already a good muon discriminator.

The OPAL detector at LEP has been extensively described 
elsewhere~\cite{opal1,opal2}. 
The event generator JETSET 7.4~\cite{jetset} is used to simulate 
$\ee$ annihilation events ($\ee\to\qqbar$), including the fragmentation
of quarks into hadrons measured in the detector. The fragmentation model
has to be tuned empirically to match kinematic distributions as 
measured with the detector~\cite{tune}. The response of the detector
is also simulated.
A data set of simulated $\ee$ collisions at a centre-of-mass energy 
$\sqrt{s} = m_{\rm Z^0}$ is used for the charm identification problem.
A second Monte Carlo data set at $\sqrt{s}=189$~GeV
has been used for the muon identification problem.

\section{Problem 1: Charm quark tagging}
The Monte Carlo events had to fulfil  
preselection cuts which ensure that an event is well reconstructed 
in the detector.
These cuts result in a first overall reduction of the 
reconstruction efficiency. 
The input variables for the machine-learning algorithms
were chosen from a larger dataset of 27 variables, 
containing various jet shape variables, e.g. Fox-Wolfram moments 
and eigenvalues of the sphericity tensor, plus several secondary 
vertex variables and lepton information. 
A jet finding algorithm~\cite{cone} clusters the tracks and the calorimeter
clusters into jets. 
Only the highest energy jet per hemisphere is used.
The variables containing information about high transverse momentum leptons
were removed in order to avoid a high correlation of this charm tag with the 
charm tag using leptons. 
The 14 variables with the largest influence on the performance 
of an ANN (27-27-1 architecture), trained to classify charm versus 
not-charm, were picked from the larger set.  
The variable selection method used is equivalent 
to a method which selects the variables with the largest
connecting weights between 
the input and the first layer. This method
has been shown to perform a fairly good variable selection~\cite{varsel}.
It would be interesting to try Hessian based selection methods in 
comparison~\cite{mueller1}. A ANN classifier with a 14-14-1 architecture
was found to have the best performance. More complex architectures did not 
improve the classification.

\begin{figure}[bt]
\epsfig{file=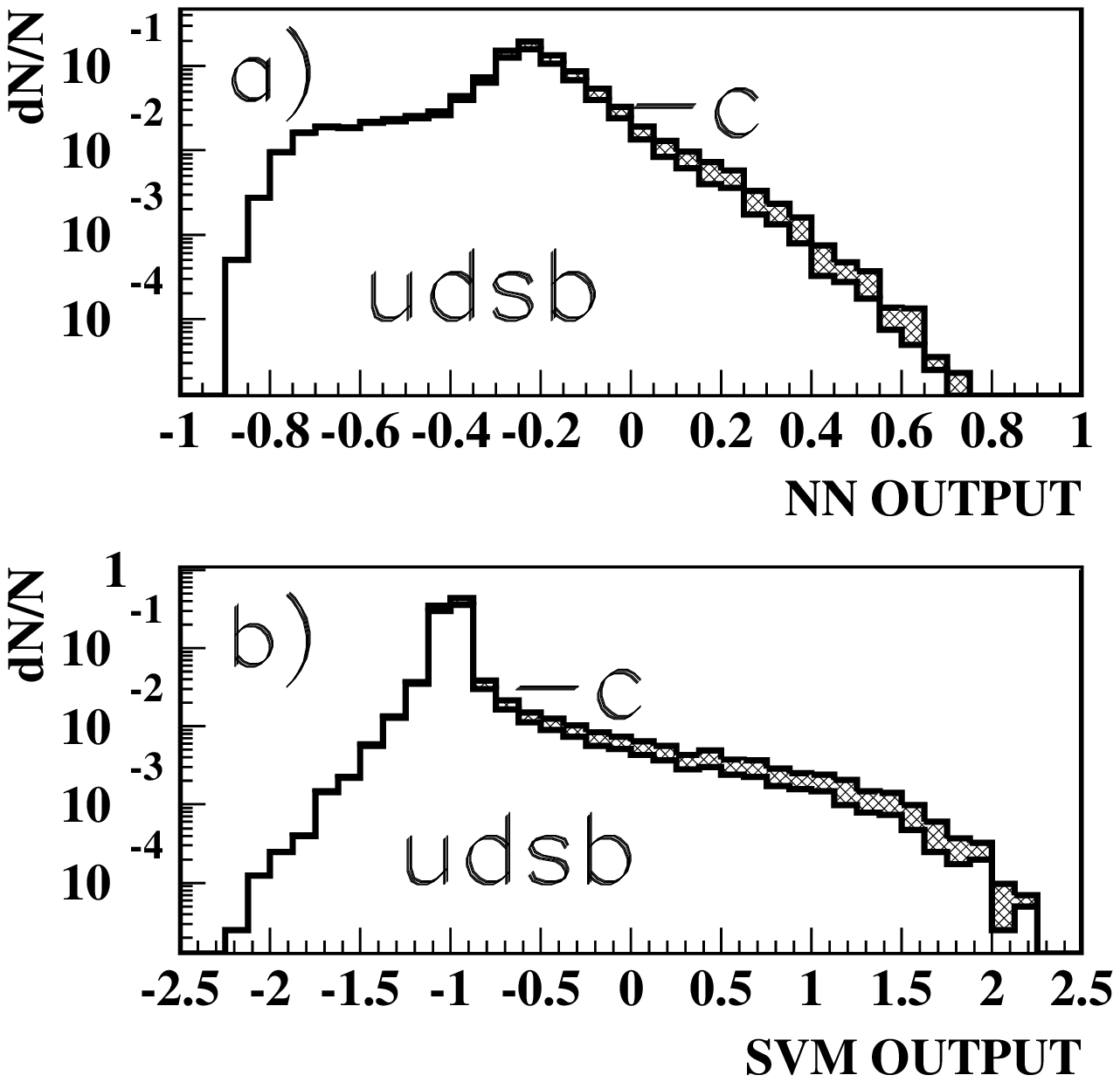,height=.5\textwidth}
\hfill
\epsfig{file=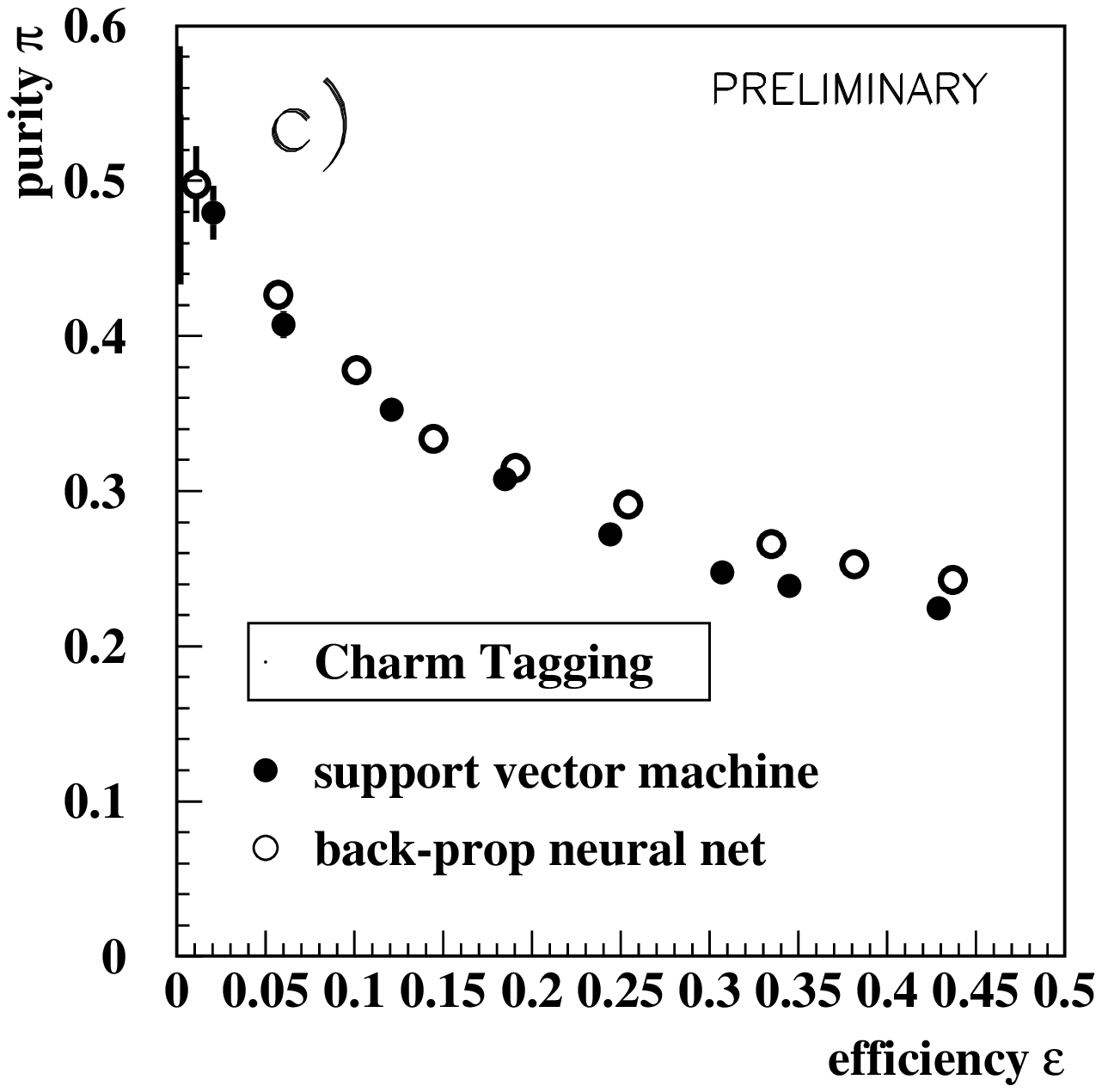,height=.5\textwidth}
\caption{The output distributions of a) the ANN and b) the SVM.
c) Purity versus efficiency for both charm tags. Statistical errors are shown.}
\label{fig:ctag}
\end{figure}

At generator level five different quark types are distinguished. 
Due to their very similar decay and jet properties, Monte Carlo events 
coming from u,d and s quarks are put into one single class (uds). 
The data set consists of 10$^5$ hemispheres per uds, c and b class. 
However, the efficiencies and purities are calculated assuming a mixture of
quark flavours according to the Standard Model prediction.
This set is divided into training, validation and test sets of equal size. 
The learning machines were trained on equal number of events 
from the three classes. The supervision during the learning phase 
consisted of a charm versus not-charm label (udsb), 
thus distinguishing only two classes.

The outputs of both learning machines are shown in Fig.~\ref{fig:ctag}. 
The two classes c and udsb are separated by
requiring a certain value for the output. This defines
the efficiency and purity of the tagged sample. 
The purity $\pi$ as a function of the efficiency $\varepsilon$ 
for the two charm tags are shown in 
Figure~\ref{fig:ctag} with the statistical errors.
The performance of the SVM is comparable to the 
performance of the ANN with a slightly higher purity $\pi$ for
the ANN at larger efficiencies $\varepsilon$.

\section{Problem 2: Muon identification}

The muon candidates are preselected by requiring a minimum track momentum of 
2 GeV and by choosing the best match in the event between muon chamber 
track segment and the extrapolated central detector track.
Ten discriminating variables containing  muon matching, 
hadron calorimeter and central detector information on the
specific energy loss, ${\rm d}E/{\rm d}x$, of charged particles were 
chosen from a larger set of variables.

\begin{figure}[bt]
\begin{center}
\epsfig{file=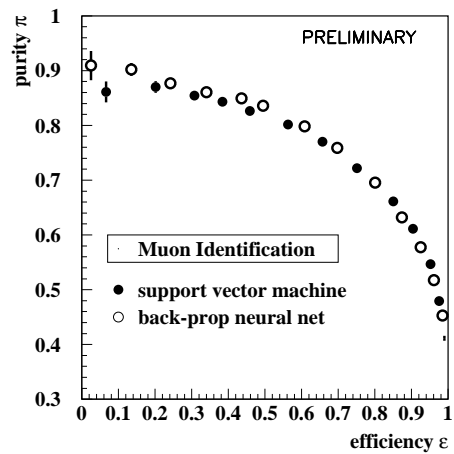,height=.5\textwidth}
\end{center}
\caption{The separation performance on identifying muons shown for a Neural Net trained with back-propagation and a Support Vector Machine. Statistical errors are shown}\label{fig:muonid}
\end{figure}

The Monte Carlo data set consists of $3 \cdot 10^4$ muons and  $3 \cdot 10^4$
 fake muon candidates. 
This set was divided into training, validation and test sets of
equal size. 
After training, the tag is defined by requiring a certain value for 
the output of the learning machines.
The resulting purity as a function of efficiency for muon identification is 
shown in Fig.~\ref{fig:muonid}. 
For high efficiency the performance of the SVM is very similar to the ANN.

\section{Conclusion}
We have compared the performance of Support Vector Machines and Artificial Neural Networks in the classification of two distinctly different problems in high energy collider physics: charm-tagging and muon identification. The constructed SVM and ANN classifiers give consistent results for efficiencies and purities of tagged samples.

\section*{Acknowledgements}
This work has been supported by the Deutsche Forschungsgemeinschaft 
with the grants SO 404/1-1, JA 379/5-2, JA 379/7-1.
We also would like to thank Frank Fiedler and Andreas Sittler of DESY for 
their help with the preparation of  the OPAL data. We thank Gunnar R\"atsch 
and Sebastian Mika of GMD-FIRST for helping to analyse the data.

\end{document}